\documentclass[aps,prb,twocolumn,shortbibliography,superscriptaddress]{revtex4-2}
\usepackage{epsfig}
\usepackage{epstopdf}
\usepackage{amsmath}
\usepackage{amsfonts}
\usepackage{amssymb}
\usepackage{hyperref}
\usepackage{bm}
\usepackage{makecell}
\usepackage{rotating}
\usepackage{hyperref}
\usepackage{caption}
\usepackage{subcaption}
\usepackage{tikz,xcolor}
\usepackage{graphicx}
\usepackage{dcolumn}
\usepackage{bm}
\usepackage{color}

\begin{document}

\title{\textbf{Non-trivial topological phases in transition metal rich half-Heusler Oxides}}

\author{Bhautik R Dhori}
\affiliation{Department of Physics, Faculty of Science, \\ The Maharaja Sayajirao University of Baroda, Vadodara, Gujarat-390 002, India}

\author{Raghottam M. Sattigeri}
\affiliation{International Research Centre MagTop, Institute of Physics, \\ Polish Academy of Sciences, Aleja Lotnik\'ow 32/46, PL-02668 Warsaw, Poland}

\author{Prafulla K Jha}
\email{prafullaj@yahoo.com}
\affiliation{Department of Physics, Faculty of Science, \\ The Maharaja Sayajirao University of Baroda, Vadodara, Gujarat-390 002, India}

\begin{abstract}
Topological Insulators with gapless surface states and insulating bulk in non-centrosymmetric cubic systems have been extensively explored following the discovery of two-dimensional quantum spin hall effect in zincblende HgTe. In such systems the negative band inversion strength E$_{BIS}$ ($=$ E$_{\Gamma_6} -$ E$_{\Gamma_8} <$ 0) governs the robustness of the non-trivial topological states at ambient conditions. Hence, realizing large negative values of E$_{BIS}$ has been a guiding motivation of several investigations reported in literature. Here, we present a material design approach which can be employed to realize large negative values of E$_{BIS}$ in cubic materials such as half-Heusler (HH) oxides with 18 valence electron configurations. We explore 27 HH oxides of the form ABO (A = Li, K, Rb; B = Cu, Ag, Au) in $\alpha$-, $\beta$-, and $\gamma$-phase (by placing transition metal atom at different Wyckoff positions) for their non-trivial topological phase. Off these three phases, we found that, the $\alpha$-phase of nine HH oxides (wherein the transition metal atoms occupy 4a Wyckoff positions in the crystal structure) is the most promising with non-trivial topological phase which is governed by the mass-darwin relativistic effects enhancing E$_{BIS}$. Whereas the other phases were found to be either trivial semiconductors or semimetals or metals and most of them being dynamically unstable. We focus on RbAuO in $\alpha$-phase with E$_{BIS}$ of $–$ 1.29 eV and the effect of strain fields on the topological surface states of this compound. We conclude that the $\alpha$-phase of HH oxide presented here can be synthesized experimentally for diverse room temperature applications in spintronics and nanoelectronics.
\end{abstract}

\maketitle

\section{Introduction}

Topological quantum materials have generated tremendous excitement in the condensed matter community over the recent years. Materials such as topological insulators (TI) \cite{1,2}, topological crystalline insulators \cite{3,4}, and kondo insulators \cite{5,6} have diverse and potential applications in spintronics, quantum computations etc. Cubic systems like HgTe with inverted band order are known to host non-trivial TI behaviour upon breaking the cubic symmetry / on creating quantum well structures etc \cite{7}. This was followed by experimental efforts which eventually lead to the confirmation of non-trivial topological phases and phase transitions in matter \cite{8}. From theoretical perspective such as, ab-initio studies, it is essential that the band order predicted in cubic structures matches those observed experimentally. In this regard, the standard generalised gradient approximations (GGA) is not accurate and reliable \cite{9}. Hence, computationally expensive methods such as, MBJGGA, SCAN etc. have been implemented which successfully predict the inverted band orders in tune with the experimental observations \cite{9,10,11}. Standard GGA approach typically fails to predict accurate inverted band orders due to band inversion strengths (defined as, E$_{BIS} =$ E$_{\Gamma_6} -$ E$_{\Gamma_8}$) of the order of meV in cubic systems. Hence, one can expect reliable results from standard GGA if the E$_{BIS}$ is large enough i.e., of the order of eV \cite{12}.

A ternary class of materials having ABO stoichiometric composition called half-Heusler (HH) oxides are a prime candidate for TI owing to their excellent thermoelectric \cite{13,14}, superconducting \cite{15,16}, and semimetallic properties \cite{13,14,15,16,17,18,31,32,33}. The non-centrosymmetric and face centred cubic crystal structure of a typical HH compound PQR constitutes P$^+$ substitution in a QR$^-$ zincblende sublattice \cite{23}. Hence, ABO HH oxides can be realised in three different phases ($\alpha$-, $\beta$-, and $\gamma$-) owing to the presence of three non-equivalent atomic arrangements within the unit cell \cite{24}. These three distinct phases correspond to C1$_b$ structure \cite{25}. Such compounds are unique for example, the atomic arrangement of non-magnetic elements can lead to a magnetic ordering in the compound \cite{19}. This implies that not only the stoichiometry but also the atomic arrangements play a vital role as evident from the literature.\cite{20,21}. 

This motivated us to explore relativistic effects such as orbital inversion and band splitting in HH oxides \cite{34}. Using E$_{BIS}$ as a critical parameter Sawai et.al., predicted a series of III-VIII-VA HH oxides with some of them hosting non-trivial topological nature under ambient conditions \cite{12}. It was proposed that inclusion of transition metals and lighter anions such as O, S and Cl may lead to large non-trivial energy gap in HH compounds \cite{20}. Shi-Yuan Lin et.al., predicted topological insulating phase in more than 2000 HH oxides based on E$_{BIS}$ \cite{20}. However, a systematic study of such HH oxides is scarce. This motivated us to thoroughly investigate ABO (A= Li, K, Rb; B= Cu, Ag, Au) HH oxides for SOC-induced topological phase.

In the present work, using ab-initio methods we have investigated the low-energy electronic structure and topological insulating phase in transition metal-rich HH oxides. We report 27 HH oxides with transition metal atoms placed at different Wyckoff positions in the unit cell. Off these, the $\alpha$-phase exhibits semimetallic electronic structure in non-relativistic regime similar to HgTe \cite{28}. These HH oxides are also analogous to well-explored binary system HgTe \cite{22} which is semimetallic in non-relativistic and relativistic regime but hosts non-trivial topological properties under dimensional confinement or uniaxial strain fields. However, unlike HgTe the relativistic regime opens up large global gap in the HH oxides with potential to host non-trivial topological phase at ambient conditions without dimensional confinement or uniaxial strain fields.

The proposed materials have large E$_{BIS}$ which implies that these compounds are excellent candidates facilitating experimental validation of non-trivial Dirac dispersions at room temperature and ambient pressure. It also known that the presence of oxygen intrinsically prevents bulk oxidations on the surface, which makes these compounds robust in addition to topological protection \cite{r6}. In addition to this, the dynamical and thermodynamical stability give directions on experimental realization of HH oxides in the $\alpha$-phase \cite{28,29}. Thus, indicating that ABO HH oxides are a potential candidate for room temperature and ambient pressure applications in the field of nanoelectronics and spintronics.

\section{Methodology}

HH oxides of the form ABO were investigated using density functional theory based \textit{first-principles} calculations as implemented in Quantum ESPRESSO \cite{qe}. Norm conserving pseudopotential based on Martins-Troullier method with Perdew-Burke-Ernzerhof type exchange-correlation functional under generalized gradient approximation was used to obtain the ground state properties of the materials \cite{35}. The pseudopotentials account for the 4s$^1$ orbitals of alkali metals, 4s$^2$3d$^9$ orbitals of transition metals, and 2s$^2$2p$^4$ orbitals of O. For SOC calculations we included the effects arising from the core valence electrons using full relativistic pseudopotential based on the projector augmented wave (PAW) method. Uniform optimized computational parameters were used for all the calculations. The kinetic energy cut-off was 90 Ry with the convergence threshold criteria of at least $<$10$^{-8}$ Ry. A Monkhorst Pack \cite{36} grid with 10 $\times$ 10 $\times$ 10 \textit{k}-points was used to map the momentum space in all the calculations. The dynamic stability of the proposed oxides are analysed in term of the phonon dispersion curves obtained using Density Functional Perturbation Theory with a \textit{q}-mesh of at least 3 $\times$ 3 $\times$ 3 for all the systems. We performed ab-initio molecular dynamics (AIMD) simulations on 2 $\times$ 2 $\times$ 2 super-cell of $\alpha$-phase for 3 picoseconds (3000 femtoseconds) time step with the temperature of thermostat set to 300 K. For topological properties, we employed maximally localized Wannier functions to build the exact tight binding Hamiltonian model for HH oxides as implemented in the wannier90 code \cite{37,38}. We then used WannierTools code to compute, $\mathbb{Z}_2$ invariants, ARPES-like surface states, and slab band structures \cite{39}.
 
\section{Results and Discussion}

\subsection{Crystal structure and stability}

\begin{figure}[ht!]
 	\centering
 	\includegraphics[width = 1\linewidth]{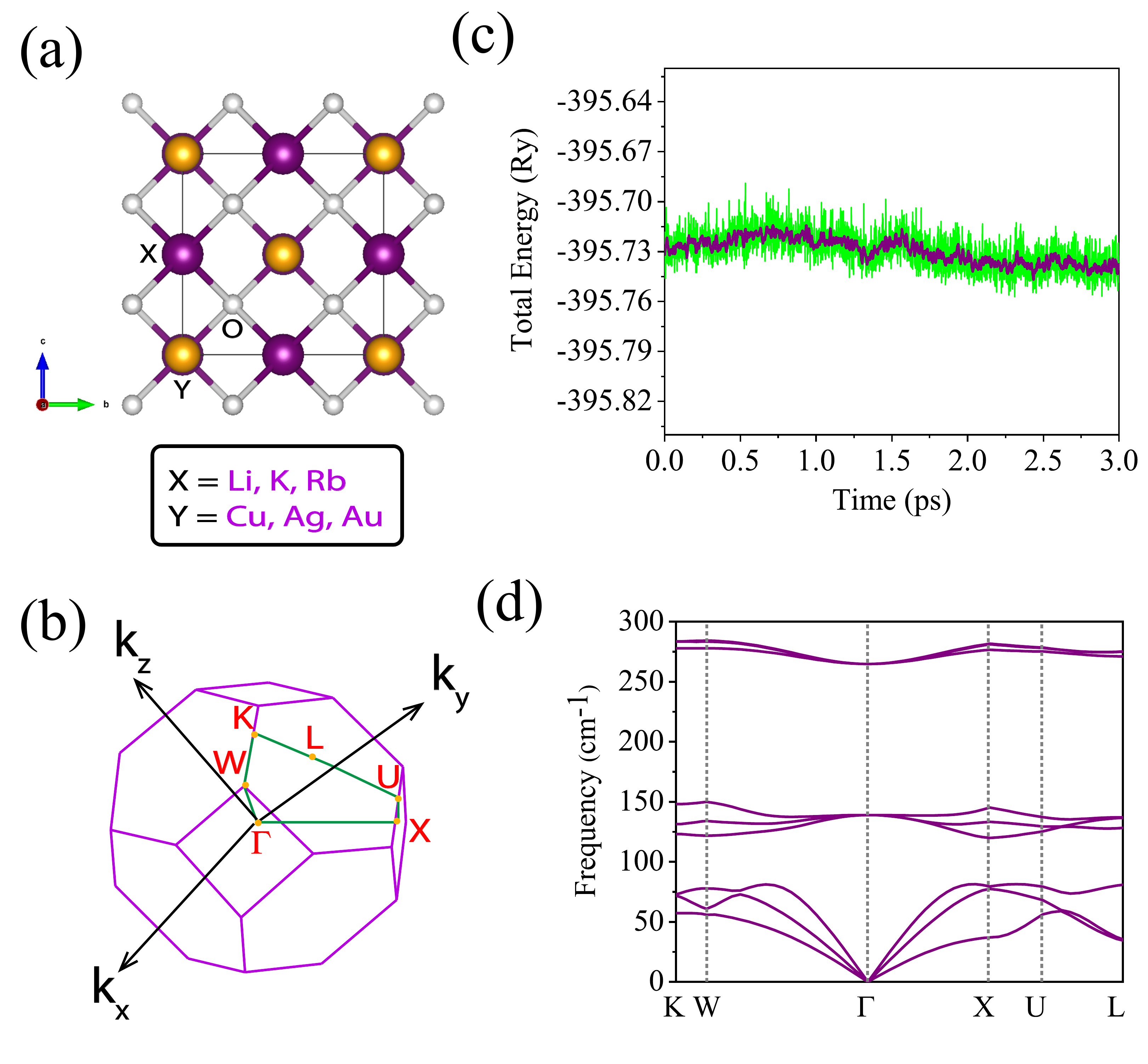}
 	\caption{\label{crystal} (a) Crystal structure of the proposed half-Heusler oxides in $\alpha$-phase. (b) Irreducible Brillouin zone corresponding to the F$\overline{4}$3m space group. (c) Ab-initio molecular dynamics simulation for 3 ps at 300 K thermostat temperature for $\alpha$-RbAuO system indicating energetic stability of the structure at room temperature. (d) Phonon dispersion curves of $\alpha$-RbAuO indicating absence of imaginary modes.}
\end{figure}

The ground state of the proposed ABO HH oxides exhibits face centered cubic structure governed by F$\overline{4}$3m space group emerging from the $\overline{4}$3m point group. The A, B, and O atoms occupy 4b, 4a, and 4c Wyckoff positions respectively which corresponds to the $\alpha$-phase. In this phase the elements at A site are alkali metal from group I(A) and elements at B site are transition metals from group I(B). We have also explored the $\beta$- and $\gamma$-phase by exchanging the Wyckoff positions of A, B and O sites in the ABO HH oxides (details in SUPPLEMENTARY MATERIAL (SM)). Figure \ref{crystal}(a) and (b) presents the conventional crystal structure and irreducible Brillouin zone of the proposed HH oxides. In such structures the Pauling electronegativity of constituent elements plays a vital role as a result, the electropositive element at A shares \lq{\textit{n}}\rq electrons with the electronegative element at B and O. This leads to HgTe-like zincblende lattice is formed by B and O atom BO$^{n-}$ and an empty void within the zincblende structure is stuffed with A atoms A$^{n+}$. The crystal structure of the proposed HH oxides were obtained from literature \cite{r1} followed by optimization of the lattice parameters by performing a proper convergence test. We have summarized the optimized lattice parameters in Tab. \ref{summary} for $\alpha$-, $\beta$- and the $\gamma$-phase of HH oxides.

Since these oxides have not yet been synthesized experimentally and due to scare data on their thermodynamic stability, we performed AIMD simulations at room temperature (300 K) for time period of 3 picoseconds (3000 femtoseconds). The AIMD results for RbAuO in $\alpha$-phase are presented in Fig. \ref{crystal}(c). It is evident that the structure thermalizes without structural deformations at room temperature with mild variations in the total energy. AIMD for other compounds in $\alpha$-phase are presented in SM Fig. 1. The AIMD results give insight into the feasibility of material synthesis at room temperature. Apart from structural stability established from the AIMD studies, we also report the phonon dispersion curves of all the phases using DFPT (SM Fig. 2,22,24). In Fig. \ref{crystal}(d) we present the phonon dispersion curves for RbAuO which indicates the absence of imaginary modes implying dynamical stability of the structure in $\alpha$-phase. 

\subsection{Electronic band structure}

The 9 HH oxides in $\alpha$-phase host semimetallic electronic structure in non-relativistic regime with degenerate valence and conduction bands at the center of the BZ. However, in the relativistic regime the bands are non-degenerate with inverted orbital character across the Fermi energy being retained at high symmetry point $\Gamma$ while being absent at other time reversal invariant momenta points in the BZ due to the lack of spatial inversion symmetry.

\begin{figure}[ht!]
	\centering
	\includegraphics[width = 1\linewidth]{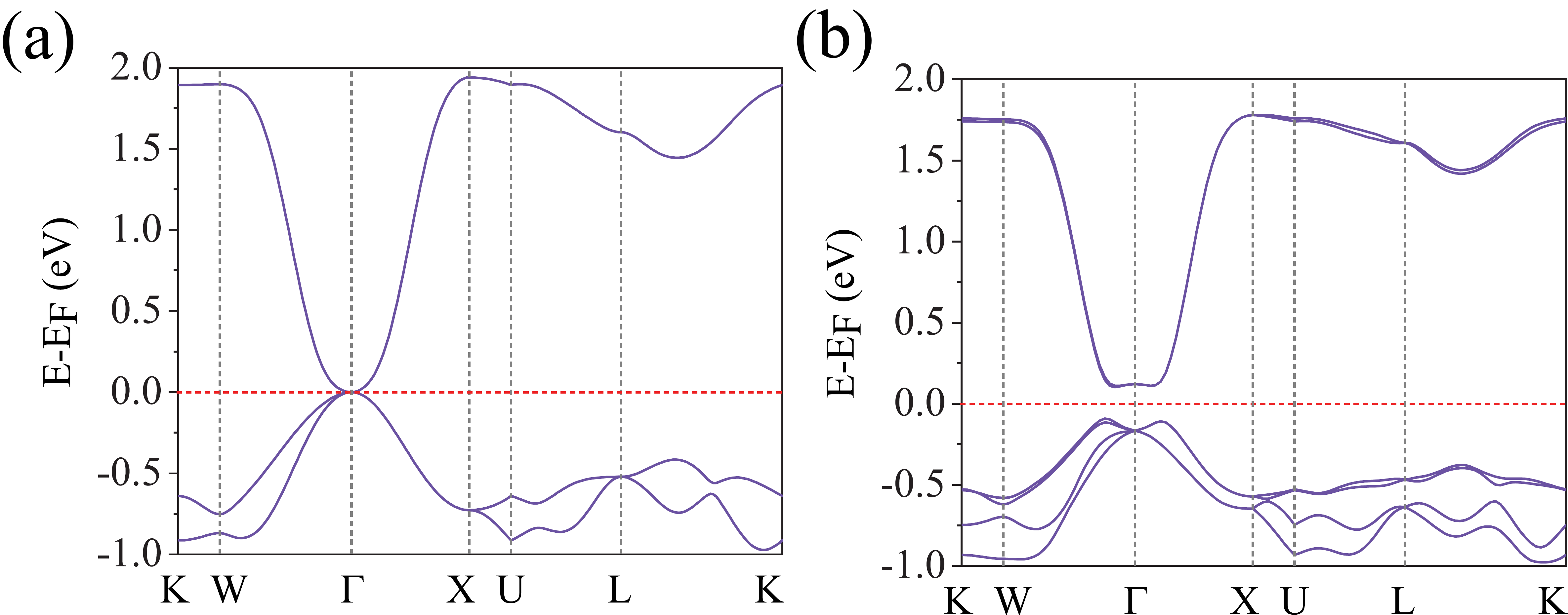}
	\caption{\label{bs} Electronic band structure of RbAuO in $\alpha$ phase (a) without and (b) with spin-orbit coupling indicating the relativistic splitting due to strong spin-orbit effects.}
\end{figure}

Hence, under ambient conditions the relativistic effects originating from the strong SOC in these HH oxides gives rise to non-trivial topological character without the influence of external strain/pressure. This eliminates the use of techniques such as, breaking crystal symmetry or dimensional confinement which is otherwise required to realise non-trivial topology in HgTe and other similar systems. The band topology of these HH oxides is governed by the $\Gamma_{7}$ (j = 1/2; 5/2) and $\Gamma_{8}$ (j = 3/2; 5/2) states in conduction band minima and valence band maxima respectively (which are composed of the -\textit{p}, -\textit{d} orbitals), and $\Gamma_{6}$ (j = 1/2; 5/2) states below the valence band maxima (which is composed of -\textit{s}, -\textit{d} orbitals). The orbital inversion mechanism involves $\Gamma_{7}$, $\Gamma_{8}$ and $\Gamma_{6}$ states with the band inversion strength defined as, E$_{BIS} = $ E$_{\Gamma_{6}} -$ E$_{\Gamma_{8}}$. The electronegativity of lighter anion and strong SOC of heavier transition metals arising from the core electrons results in this intrinsic band inversion. For E$_{BIS} <$ 0 the HH oxides exhibit non-trivial topological phase, while a trivial topological phase is observed when E$_{BIS} > $ 0. All HH oxides in $\alpha$-phase host inverted band characteristic under equilibrium conditions with $\Gamma_{6}$ states below the $\Gamma_{8}$ states. The calculated value of E$_{BIS}$ and the SOC induced global gap for all these HH oxides are tabulated in Tab. \ref{summary}, which clearly demonstrates non-trivial topological character (further confirmed by computing $\mathbb{Z}_2$ invariant). In all these 9 HH oxides in $\alpha$-phase, we notice indistinguishable electronic band structures (as shown in SM Fig. 3) and with topological band ordering (as shown in SM Fig. 4). Off the 9 HH oxides, 7 (LiCuO, LiAgO, LiAuO, KCuO, KAuO, RbCuO, RbAuo) exhibit a Mexican hat-like band dispersion in their conduction band minima and valence band maxima due to non-trivial orbital inversions resulting in higher values of E$_{BIS}$ as compared to the global energy gap. 

\begin{figure}[h]
	\centering
	\includegraphics[width = 1\linewidth]{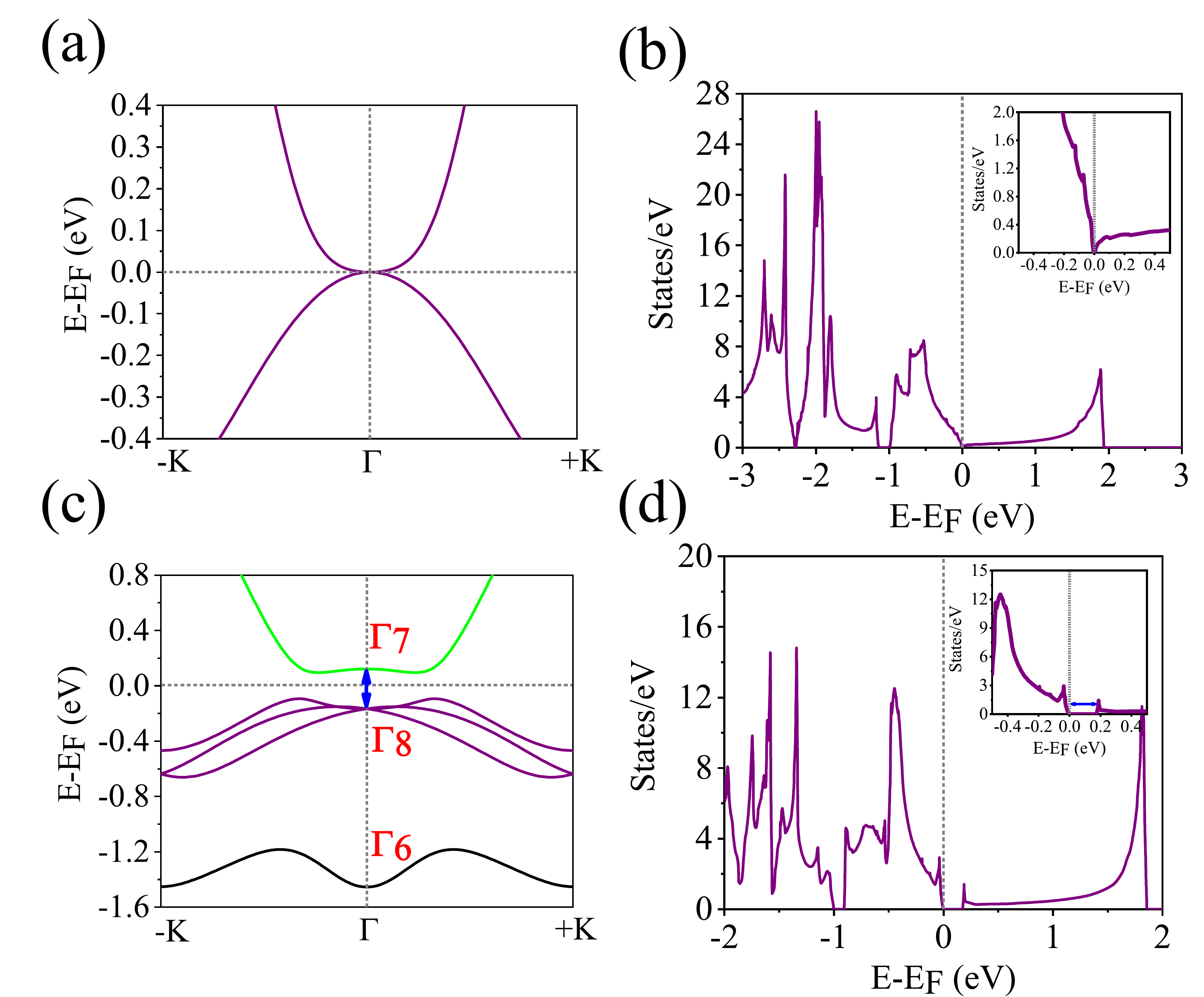}
	\caption{\label{rbauo-bands} (a-b) The degenerate electronic states at $\Gamma$ point in absence of relativistic effects and the density of states indicating the semi-metallic nature. (c-d) Non-trivial global gap evident in the electronic structure and density of states indicating insulating characters.}
\end{figure}

\begin{figure*}[ht!]
 	\centering
 	\includegraphics[width = 1\linewidth]{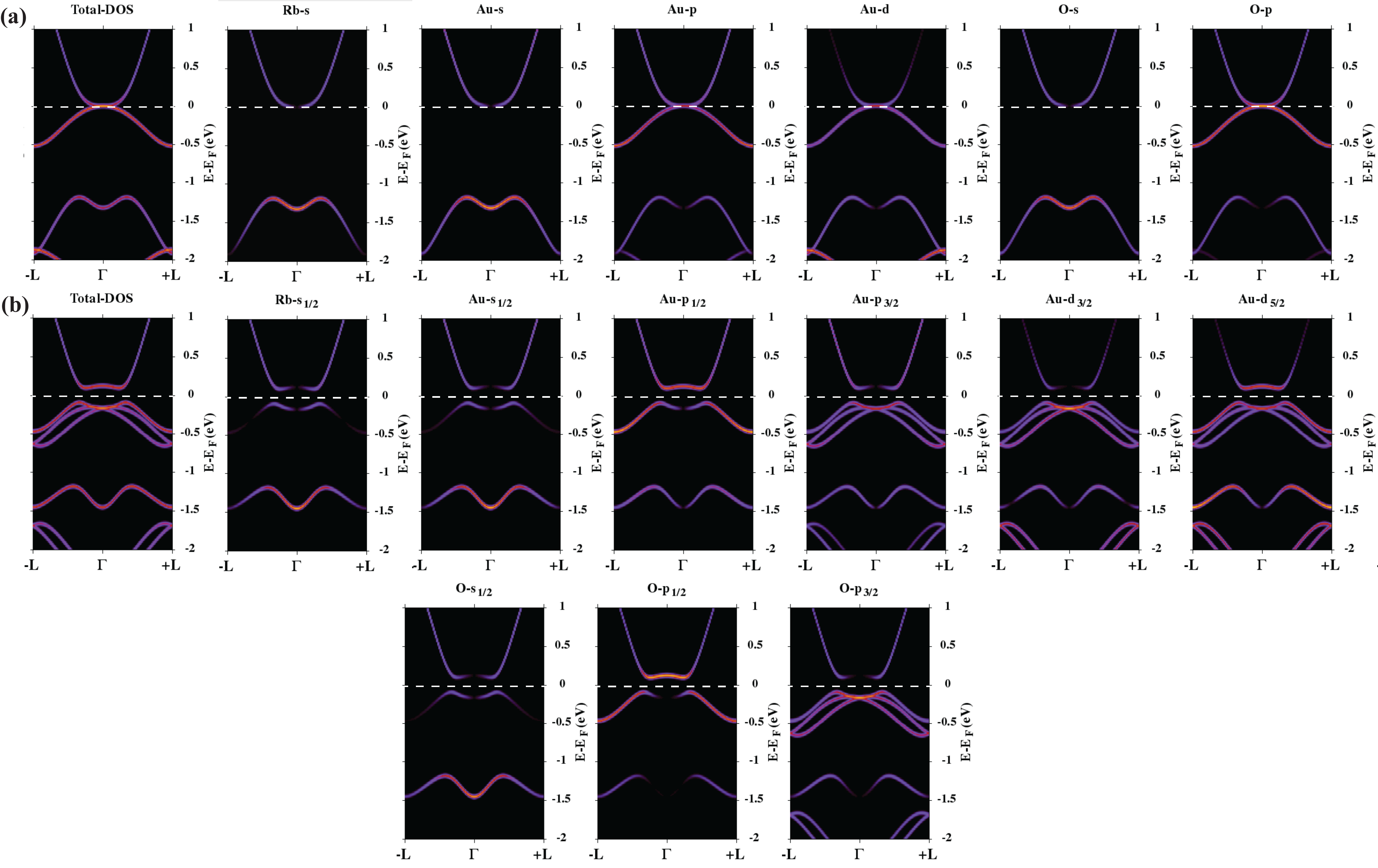}
 	\caption{\label{fat-bands} Atomic and orbital projected band structures for RbAuO. (a) Total density of states projected onto the electronic band structures in low energy regime without SOC alongside atomic and orbital projections. (b) Total density of states projected onto the electronic band structures in low energy regime with SOC alongside atomic and orbital projections.}
\end{figure*}

Other HH oxides with the transition metals placed at different Wyckoff positions leading to the $\beta$- and $\gamma$-phase were investigated. We do not discuss their topological properties since these compounds are either trivial insulators, semimetals or metals with all of them being dynamically unstable as evident from the phonon dispersions presented in Fig. 22 and 24 in SM. Hence, in the following sections, we primarily focus only on the dynamically stable compound RbAuO in $\alpha$-phase for its electronic and topological properties owing to strong SOC and large E$_{BIS}$.

\subsection{The case of RbAuO}

\subsubsection{Electronic and Topological properties}

With the thermodynamic and dynamical stability of RbAuO presented in Fig. \ref{crystal}(c,d) we report the electronic and topological properties under equilibrium conditions presented in Fig. \ref{rbauo-bands}. The optimized lattice parameters for RbAuO are presented in Tab. \ref{summary}. The electronic band structures in absence and presence of the relativistic effects are presented in Fig. \ref{bs} at equilibrium conditions. In non-relativistic regime the doubly degenerate states at $\Gamma$ point are protected by the cubic symmetry of the system. However, in the relativistic regime the conduction and valence bands are well resolved with a global gap of 190 meV in the entire BZ with Fermi energy in between. Since the DOS of RbAuO, without the inclusion of SOC, is zero at the Dirac points (as shown in Fig. \ref{rbauo-bands} (b)). In the presence of SOC, the gap appears and the DOS are reducing near the Fermi level (Fig. \ref{rbauo-bands} (d)). The low energy excitations and orbital band inversions are analyzed near the Fermi energy in terms of the atomic and orbital projected electronic structures wherein for TIs without inversion symmetry, the band inversion occurs at a specific time reversal invariant momenta in the reciprocal space.

The atomic and orbital projected electronic structures in the non-relativistic and relativistic regime are presented in Fig. \ref{fat-bands}. In the non-relativistic regime the crystal field gives rise to intrinsically inverted band order with semimetallic character being evident from the degenerate conduction and valence bands. This degeneracy originates from the -\textit{p} and -\textit{d} orbitals of Au and O atoms in the unit cell while the lighter -\textit{s} orbitals are pushed deeper below the Fermi energy which is a clear indication of the inverted bulk band order making the system non-trivial.

However, in the relativistic regime, the strong SOC effects enhance the inverted orbital order which already existed due to the crystal field. The -\textit{p}$_{1/2}$, -\textit{p}$_{3/2}$, -\textit{d}$_{3/2}$, and -\textit{d}$_{5/2}$ orbitals originating from the Au and O atoms in the unit cell dominate the conduction band minima and the valence band maxima. While the -\textit{s}$_{1/2}$ orbitals originating from Rb, Au, and O get pushed even farther from the Fermi energy (from -1.4 eV to -1.5 eV) as compared to the non-relativistic regime with some contribution from the -\textit{d}$_{5/2}$ orbitals of Au atoms. This inverted order is also enhanced due to the mass-darwin relativistic effects in the $\alpha$-phase of HH oxides. This leads to high E$_{BIS} = -$1.29 eV with well resolved global energy gap of 190 meV in the entire BZ which is quite high as compared to several HH compounds \cite{r2,r3}. Hence, the intrinsically inverted orbital order indicates that RbAuO is non-trivial.

\begin{figure}[h]
 	\centering
 	\includegraphics[width = 1\linewidth]{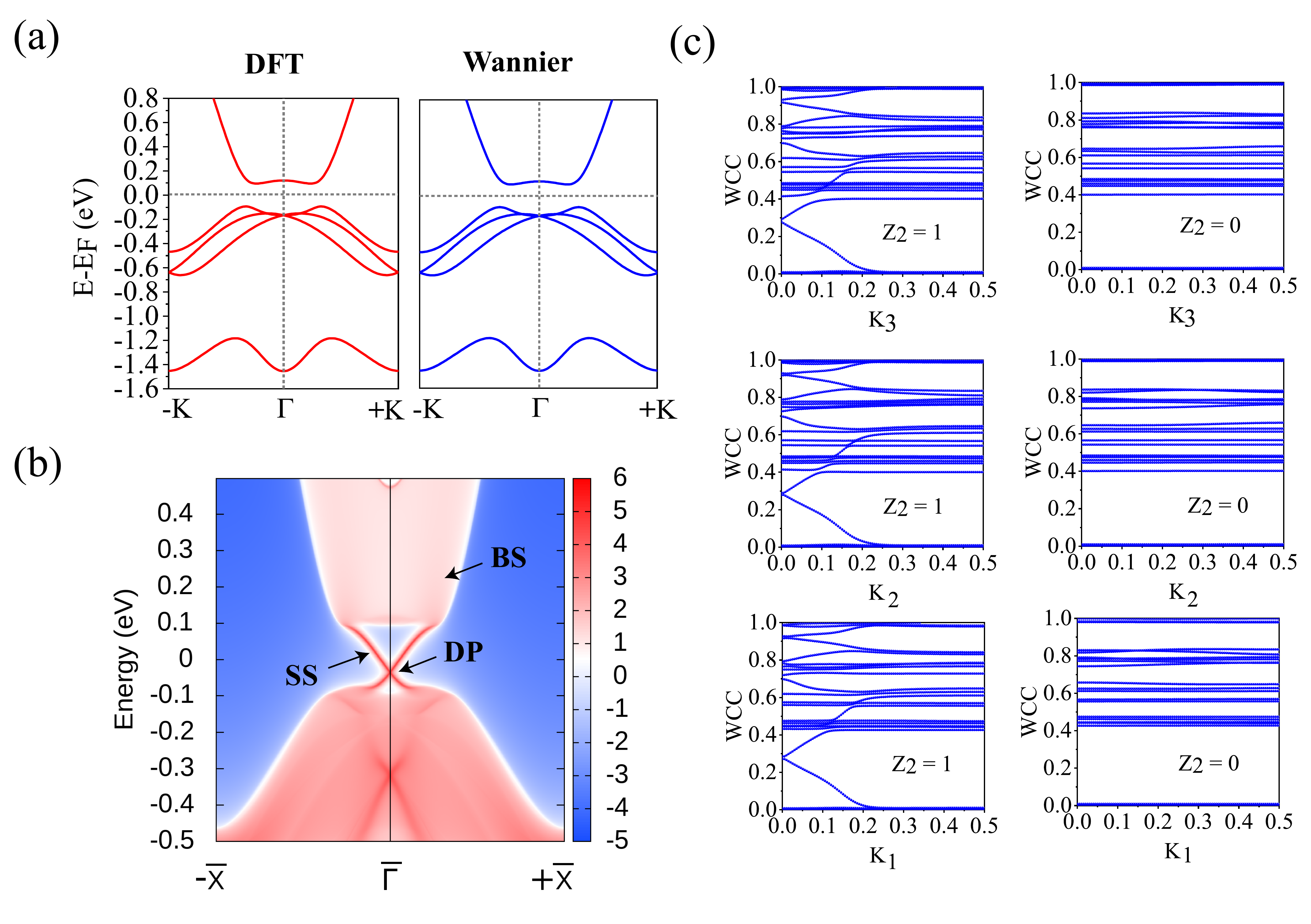}
 	\caption{\label{ss} (a) Comparison of electronic band structure from DFT and wannierization (b) Computational surface states of the RbAuO oxides demonstrate conducting surfaces (c) evolution of Wannier charge centers}
\end{figure}

This type of non-trivial inverted band orders involving -\textit{s}, -\textit{p}, and -\textit{d} orbitals has also been observed in other HH compounds such as ScPtBi, LuPdBi, and other C1$_{b}$ type HH oxides \cite{r4,r7}. Such inverted band order is also observed in HH oxides NaYO (Y = Ag, Au, and Cu) \cite{r4}.

The non-trivial nature of the proposed HH oxides and the corresponding band topologies are further classified using $\mathbb{Z}_2$ invariants. Since the system does not posses inversion symmetry, we compute these invariants using the Wilson loop method. 
\begin{equation}\label{eq1}
\nu_0 = (\mathbb{Z}_{2_{(k_i = 0)}} + \mathbb{Z}_{2_{(k_i = 0.5)}}) \cdot mod 2
\end{equation}

\begin{equation}\label{eq2}
\nu_i = \mathbb{Z}_{2_{(k_i = 0.5)}}
\end{equation}

This is achieved by computing the maximally localized Wannier functions to generate the exact tight-binding Hamiltonian for the system. Hence, $\mathbb{Z}_2$ invariants along six time reversal invariant planes i.e., k$_x$ = 0,$\pi$, k$_y$ = 0,$\pi$ and k$_z$ = 0,$\pi$ in the BZ are computed using equation \ref{eq1} and \ref{eq2} rather than computing the product of parity of the eigen functions \cite{r5}.

Figure \ref{ss}(a) presents the exact match between the electronic structure calculated using DFT and the electronic structure extracted from the exact tight-binding Hamiltonian generated using wannierization. The $\mathbb{Z}_2$ invariants are obtained by tracing the evolution of wannier charge centres in the momentum space presented in Fig. \ref{ss} (c) which is obtained by using the Wilson loop method. A non-trivial topology is characterized by the presence of Kramer's doublet exchange occuring during the evolution of wannier charge centers. For non-trivial systems, the evolution of wannier charge centres presented in Fig. \ref{ss}(c) odd(even) number of crossings for $k_{i = 0}$ and $k_{i = 0.5}$ respectively. Thus the $\mathbb{Z}_2$ invariants are ($\nu_0$, $\nu_1$ $\nu_2$ $\nu_3$) = (1, 000) quantifying the non-trivial nature of RbAuO. Since $\nu_0 \neq 0$ it implies that the proposed system RbAuO is a strong TI. Furthermore, using Green's function method, we calculate ARPES-like surface state spectra which is presented in Fig. \ref{ss}(b). It is evident that, the bulk global gap in the momentum space hosts a robust Dirac dispersion on the surface projected along [111] direction. Hence, RbAuO can host dissipationless transport of Fermions across the surface originating from the robust inverted band order at room temperatures owing to high E$_{BIS}$. We summarise the structural properties of other HH oxides in $\beta$- and $\gamma$-phase in Tab. \ref{summary}.

\subsubsection{Strain effects}

\begin{figure}[h]
 	\centering
 	\includegraphics[width = 1\linewidth]{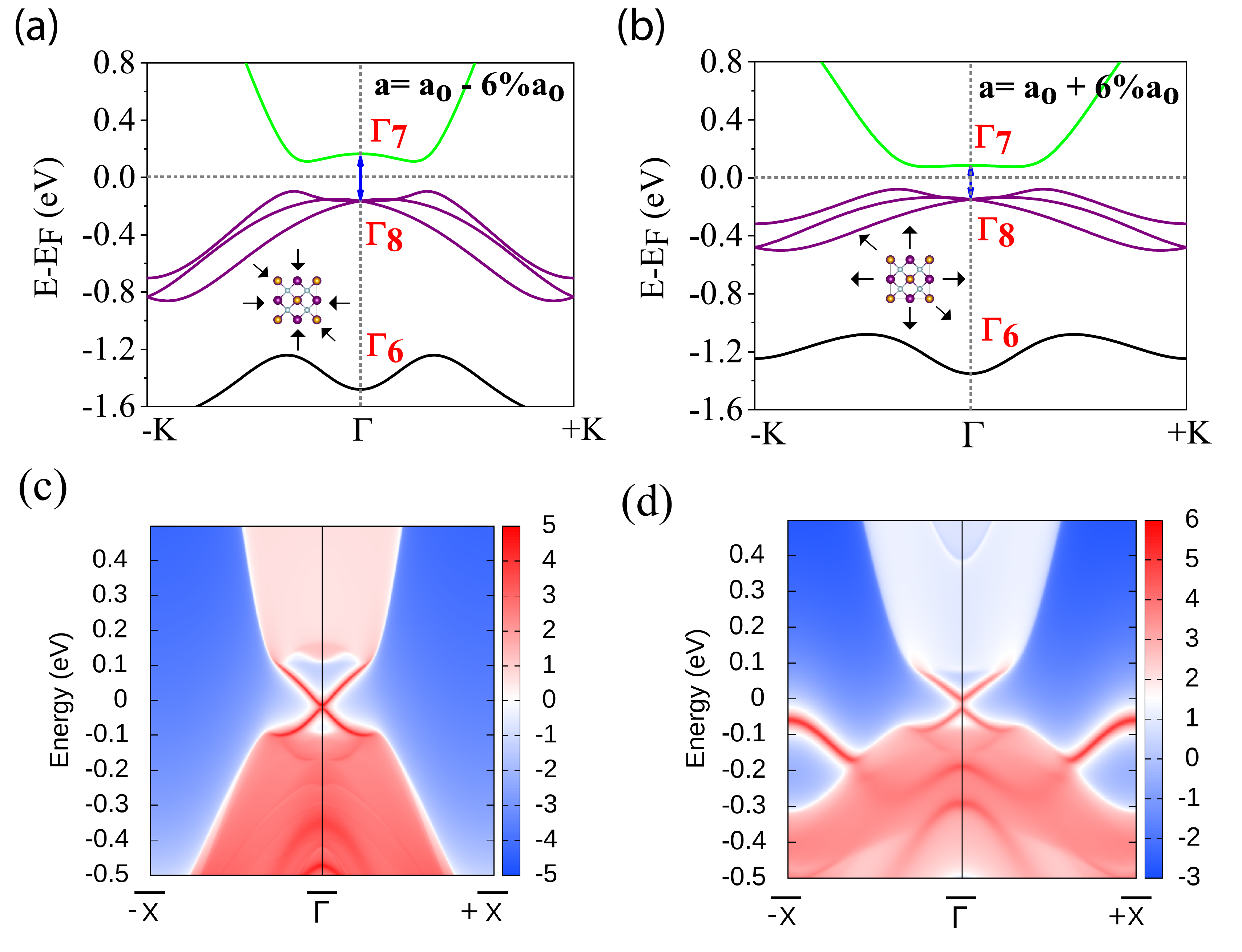}
 	\caption{\label{band} Electronic band structures with SOC at (a) a $=$ a$_0 - $ 6\% a$_0$ and (b) a $=$ a$_0 + $ 6\% a$_0$ indicating band inversion at high symmetric point $\Gamma$. ARPES-like surface state spectra for (c) compressive and (d) tensile strains at $\pm$ 6\%.}
\end{figure}

\begin{table*}[ht!]
\caption{Summary of atomic Wyckoff positions and lattice parameters of the proposed transition metal rich HH oxides in $\alpha$-, $\beta$- and $\gamma$-phase. Along with the energy band gap (E$_g$), band inversion strength (E$_{BIS}$) and the $\mathbb{Z}_2$ invariant for the $\alpha$-phase. Note that, the symbols $^*$, $^\dagger$ and, $^\#$ correspond to atoms presented in the respective columns above.}\label{summary}
    \centering
    \resizebox{\linewidth}{!}
    {\begin{tabular}{|c|c|c|c|c|c|c|c|c|c|}
    \hline
    Compounds $\xrightarrow{}$ & Li$^*$Cu$^\dagger$O$^\#$ & Li$^*$Ag$^\dagger$O$^\#$ & Li$^*$Au$^\dagger$O$^\#$ & K$^*$Cu$^\dagger$O$^\#$ & K$^*$Ag$^\dagger$O$^\#$ & K$^*$Au$^\dagger$O$^\#$ & Rb$^*$Cu$^\dagger$O$^\#$ & Rb$^*$Ag$^\dagger$O$^\#$ & Rb$^*$Au$^\dagger$O$^\#$ \\
    \hline
    $\alpha$-phase & 4b$^*$4a$^\dagger$4c$^\#$ & 4b$^*$4a$^\dagger$4c$^\#$ & 4b$^*$4a$^\dagger$4c$^\#$ & 4b$^*$4a$^\dagger$4c$^\#$ & 4b$^*$4a$^\dagger$4c$^\#$ & 4b$^*$4a$^\dagger$4c$^\#$ & 4b$^*$4a$^\dagger$4c$^\#$ & 4b$^*$4a$^\dagger$4c$^\#$ & 4b$^*$4a$^\dagger$4c$^\#$ \\
    \hline
    $\beta$-phase & 4a$^*$4c$^\dagger$4b$^\#$ & 4a$^*$4c$^\dagger$4b$^\#$ & 4a$^*$4c$^\dagger$4b$^\#$ & 4a$^*$4c$^\dagger$4b$^\#$ & 4a$^*$4c$^\dagger$4b$^\#$ & 4a$^*$4c$^\dagger$4b$^\#$ & 4a$^*$4c$^\dagger$4b$^\#$ & 4a$^*$4c$^\dagger$4b$^\#$ & 4a$^*$4c$^\dagger$4b$^\#$ \\
    \hline
    $\gamma$-phase & 4c$^*$4b$^\dagger$4a$^\#$ & 4c$^*$4b$^\dagger$4a$^\#$ & 4c$^*$4b$^\dagger$4a$^\#$ & 4c$^*$4b$^\dagger$4a$^\#$ & 4c$^*$4b$^\dagger$4a$^\#$ & 4c$^*$4b$^\dagger$4a$^\#$ & 4c$^*$4b$^\dagger$4a$^\#$ & 4c$^*$4b$^\dagger$4a$^\#$ & 4c$^*$4b$^\dagger$4a$^\#$ \\
    \hline
    $\alpha$-phase, a (\AA) & 4.95 & 5.45 & 5.50 & 5.74 & 5.99 & 6.04 & 5.98 & 6.19 & 6.24 \\
    \hline
    $\beta$-phase, a (\AA) & 5.02 & 5.53 & 5.59 & 5.75 & 6.17 & 6.34 & 5.88 & 6.33 & 6.55 \\
    \hline
    $\gamma$-phase, a (\AA) & 4.91 & 5.30 & 5.39 & 5.97 & 6.24 & 6.41 & 6.18 & 6.52 & 6.72 \\
    \hline
    $\alpha$-phase, E$_g$ (meV) & 65.3 & 110.0 & 84.5 & 76.6 & 76.8 & 186.0 & 65.9 & 53.6 & 190.0 \\
    \hline
    $\alpha$-phase, E$_{BIS}$ (eV) & $-$1.25 & $-$0.97 & $-$2.37 & $-$0.72 & $-$0.19 & $-$1.47 & $-$0.59 & $-$0.08 & $-$1.29 \\
    \hline
    $\alpha$-phase, $\mathbb{Z}_2$ & 1 & 1 & 1 & 1 & 1 & 1 & 1 & 1 & 1 \\
    \hline
    \end{tabular}}
    \end{table*}

Since RbAuO exhibits high E$_{BIS}$, we further explore the robustness of the non-trivial topological nature by subjecting the system to hydrostatic strain of compressive and tensile nature retaining the cubic symmetry. We find that the electronic and topological properties of RbAuO has drast variations in the wide range of strain $\pm$6\%. In the tensile regime, the system global gap reduces with the the Dirac cone being annihilated on the surface, making the system weak TI. Whereas in the compressive regime, the Dirac cone is retained and the global gap in the momentum space gets enhanced reaching a maximum value of 230 meV. This is evident from electronic structure presented in Fig. \ref{band}(a,b). Although, the inverted orbital order is retained in both; the tensile and compressive regimes, the stark difference is due to weaker orbital hybridization's in the tensile regime as compared to the compressive regime. Hence, we show that the, non-trivial nature and Dirac dispersions along the surface would persist under compressive strains and annihilated under tensile strains. The compressive strains can be realised in experimental conditions subjecting the system to substrate effects hence indicating towards device applications for spintronics and nanoelectronics at room temperature.

\section{Conclusion}

To summarise, we propose a design approach wherein transition metals are placed at different Wyckoff positions leading to 27 transition metal rich HH oxides in $\alpha$-, $\beta$- and $\gamma$-phase. We show that, off the 27 predicted materials, 18 are dynamically unstable in the $\beta$- and $\gamma$-phase which makes $\alpha$-phase the most promising configuration of the HH oxides. However, off the 9 HH oxides in $\alpha$-phase, we have 4 systems which are dynamically unstable with thermodynamic stability being observed at room temperature. This left us with only 5 dynamically stable configurations in the $\alpha$-phase. Off these, RbAuO was found to be the most promising HH oxide owing to its high E$_{BIS} = -$ 1.29 eV which is quite high compared to other compounds with a global energy gap in the entire BZ of 190 meV under equilibrium conditions. With such strong inverted band order originating from the mass-darwin relativistic effects, we further report the non-trivial topological nature in terms of the $\mathbb{Z}_2$ invariant and ARPES-like surface states. To assess the practicality in device applications of the proposed compound, we subjected the system to tensile and compressive strains of $\pm$6\%. The results show that, under compressive regime of strain the electronic bands under go blue shift in energy with increment in the global energy gap to maximum of 230 meV. With the non-trivial inverted band order being retained which is evident from the $\mathbb{Z}_2$ invariant and the Dirac cone on the projected surface states. Whereas, in the tensile strain, the Dirac cone is anihilated with the system exhibiting weak TI nature. Thus we propose transition metal rich HH oxides in $\alpha$-phase, especially RbAuO for spintronic and nanoelectronic applications under ambient conditions and at room temperature.

\section*{Conflicts of interest}
There are no conflicts to declare.

\section*{Acknowledgement}
BRD highly appreciates the Department of Science and Technology (DST), Government of India, for the Inspire Fellowship award (No. DST/INSPIRE fellowship/2020/IF200265). RMS acknowledges the support from Foundation for Polish Science through the International Research Agendas program co-financed by the European Union within the Smart Growth Operational Programme (Grant No. MAB/2017/1) and access to the computing facilities of the Poznan Supercomputing and Networking Center Grant No. 609. 

\bibliography{main}

\providecommand{\noopsort}[1]{}\providecommand{\singleletter}[1]{#1}%
\begin{thebibliography}{44}%
\makeatletter
\providecommand \@ifxundefined [1]{%
 \@ifx{#1\undefined}
}%
\providecommand \@ifnum [1]{%
 \ifnum #1\expandafter \@firstoftwo
 \else \expandafter \@secondoftwo
 \fi
}%
\providecommand \@ifx [1]{%
 \ifx #1\expandafter \@firstoftwo
 \else \expandafter \@secondoftwo
 \fi
}%
\providecommand \natexlab [1]{#1}%
\providecommand \enquote  [1]{``#1''}%
\providecommand \bibnamefont  [1]{#1}%
\providecommand \bibfnamefont [1]{#1}%
\providecommand \citenamefont [1]{#1}%
\providecommand \href@noop [0]{\@secondoftwo}%
\providecommand \href [0]{\begingroup \@sanitize@url \@href}%
\providecommand \@href[1]{\@@startlink{#1}\@@href}%
\providecommand \@@href[1]{\endgroup#1\@@endlink}%
\providecommand \@sanitize@url [0]{\catcode `\\12\catcode `\$12\catcode
  `\&12\catcode `\#12\catcode `\^12\catcode `\_12\catcode `\%12\relax}%
\providecommand \@@startlink[1]{}%
\providecommand \@@endlink[0]{}%
\providecommand \url  [0]{\begingroup\@sanitize@url \@url }%
\providecommand \@url [1]{\endgroup\@href {#1}{\urlprefix }}%
\providecommand \urlprefix  [0]{URL }%
\providecommand \Eprint [0]{\href }%
\providecommand \doibase [0]{http://dx.doi.org/}%
\providecommand \selectlanguage [0]{\@gobble}%
\providecommand \bibinfo  [0]{\@secondoftwo}%
\providecommand \bibfield  [0]{\@secondoftwo}%
\providecommand \translation [1]{[#1]}%
\providecommand \BibitemOpen [0]{}%
\providecommand \bibitemStop [0]{}%
\providecommand \bibitemNoStop [0]{.\EOS\space}%
\providecommand \EOS [0]{\spacefactor3000\relax}%
\providecommand \BibitemShut  [1]{\csname bibitem#1\endcsname}%
\let\auto@bib@innerbib\@empty
\bibitem [{\citenamefont {Bernevig}\ and\ \citenamefont {Zhang}(2006)}]{1}%
  \BibitemOpen
  \bibfield  {author} {\bibinfo {author} {\bibfnamefont {B.~A.}\ \bibnamefont
  {Bernevig}}\ and\ \bibinfo {author} {\bibfnamefont {S.-C.}\ \bibnamefont
  {Zhang}},\ }\href@noop {} {\bibfield  {journal} {\bibinfo  {journal}
  {Physical review letters}\ }\textbf {\bibinfo {volume} {96}},\ \bibinfo
  {pages} {106802} (\bibinfo {year} {2006})}\BibitemShut {NoStop}%
\bibitem [{\citenamefont {Hasan}\ and\ \citenamefont {Kane}(2010)}]{2}%
  \BibitemOpen
  \bibfield  {author} {\bibinfo {author} {\bibfnamefont {M.~Z.}\ \bibnamefont
  {Hasan}}\ and\ \bibinfo {author} {\bibfnamefont {C.~L.}\ \bibnamefont
  {Kane}},\ }\href@noop {} {\bibfield  {journal} {\bibinfo  {journal} {Reviews
  of modern physics}\ }\textbf {\bibinfo {volume} {82}},\ \bibinfo {pages}
  {3045} (\bibinfo {year} {2010})}\BibitemShut {NoStop}%
\bibitem [{\citenamefont {Fu}(2011)}]{3}%
  \BibitemOpen
  \bibfield  {author} {\bibinfo {author} {\bibfnamefont {L.}~\bibnamefont
  {Fu}},\ }\href@noop {} {\bibfield  {journal} {\bibinfo  {journal} {Physical
  Review Letters}\ }\textbf {\bibinfo {volume} {106}},\ \bibinfo {pages}
  {106802} (\bibinfo {year} {2011})}\BibitemShut {NoStop}%
\bibitem [{\citenamefont {Tanaka}\ \emph {et~al.}(2012)\citenamefont {Tanaka},
  \citenamefont {Ren}, \citenamefont {Sato}, \citenamefont {Nakayama},
  \citenamefont {Souma}, \citenamefont {Takahashi}, \citenamefont {Segawa},\
  and\ \citenamefont {Ando}}]{4}%
  \BibitemOpen
  \bibfield  {author} {\bibinfo {author} {\bibfnamefont {Y.}~\bibnamefont
  {Tanaka}}, \bibinfo {author} {\bibfnamefont {Z.}~\bibnamefont {Ren}},
  \bibinfo {author} {\bibfnamefont {T.}~\bibnamefont {Sato}}, \bibinfo {author}
  {\bibfnamefont {K.}~\bibnamefont {Nakayama}}, \bibinfo {author}
  {\bibfnamefont {S.}~\bibnamefont {Souma}}, \bibinfo {author} {\bibfnamefont
  {T.}~\bibnamefont {Takahashi}}, \bibinfo {author} {\bibfnamefont
  {K.}~\bibnamefont {Segawa}}, \ and\ \bibinfo {author} {\bibfnamefont
  {Y.}~\bibnamefont {Ando}},\ }\href@noop {} {\bibfield  {journal} {\bibinfo
  {journal} {Nature Physics}\ }\textbf {\bibinfo {volume} {8}},\ \bibinfo
  {pages} {800} (\bibinfo {year} {2012})}\BibitemShut {NoStop}%
\bibitem [{\citenamefont {Wolgast}\ \emph {et~al.}(2013)\citenamefont
  {Wolgast}, \citenamefont {Kurdak}, \citenamefont {Sun}, \citenamefont
  {Allen}, \citenamefont {Kim},\ and\ \citenamefont {Fisk}}]{5}%
  \BibitemOpen
  \bibfield  {author} {\bibinfo {author} {\bibfnamefont {S.}~\bibnamefont
  {Wolgast}}, \bibinfo {author} {\bibfnamefont {{\c{C}}.}~\bibnamefont
  {Kurdak}}, \bibinfo {author} {\bibfnamefont {K.}~\bibnamefont {Sun}},
  \bibinfo {author} {\bibfnamefont {J.}~\bibnamefont {Allen}}, \bibinfo
  {author} {\bibfnamefont {D.-J.}\ \bibnamefont {Kim}}, \ and\ \bibinfo
  {author} {\bibfnamefont {Z.}~\bibnamefont {Fisk}},\ }\href@noop {} {\bibfield
   {journal} {\bibinfo  {journal} {Physical Review B}\ }\textbf {\bibinfo
  {volume} {88}},\ \bibinfo {pages} {180405} (\bibinfo {year}
  {2013})}\BibitemShut {NoStop}%
\bibitem [{\citenamefont {Neupane}\ \emph {et~al.}(2013)\citenamefont
  {Neupane}, \citenamefont {Alidoust}, \citenamefont {Xu}, \citenamefont
  {Kondo}, \citenamefont {Ishida}, \citenamefont {Kim}, \citenamefont {Liu},
  \citenamefont {Belopolski}, \citenamefont {Jo}, \citenamefont {Chang} \emph
  {et~al.}}]{6}%
  \BibitemOpen
  \bibfield  {author} {\bibinfo {author} {\bibfnamefont {M.}~\bibnamefont
  {Neupane}}, \bibinfo {author} {\bibfnamefont {N.}~\bibnamefont {Alidoust}},
  \bibinfo {author} {\bibfnamefont {S.}~\bibnamefont {Xu}}, \bibinfo {author}
  {\bibfnamefont {T.}~\bibnamefont {Kondo}}, \bibinfo {author} {\bibfnamefont
  {Y.}~\bibnamefont {Ishida}}, \bibinfo {author} {\bibfnamefont {D.-J.}\
  \bibnamefont {Kim}}, \bibinfo {author} {\bibfnamefont {C.}~\bibnamefont
  {Liu}}, \bibinfo {author} {\bibfnamefont {I.}~\bibnamefont {Belopolski}},
  \bibinfo {author} {\bibfnamefont {Y.}~\bibnamefont {Jo}}, \bibinfo {author}
  {\bibfnamefont {T.-R.}\ \bibnamefont {Chang}},  \emph {et~al.},\ }\href@noop
  {} {\bibfield  {journal} {\bibinfo  {journal} {Nature communications}\
  }\textbf {\bibinfo {volume} {4}},\ \bibinfo {pages} {1} (\bibinfo {year}
  {2013})}\BibitemShut {NoStop}%
\bibitem [{\citenamefont {K{\"u}fner}\ and\ \citenamefont
  {Bechstedt}(2014)}]{7}%
  \BibitemOpen
  \bibfield  {author} {\bibinfo {author} {\bibfnamefont {S.}~\bibnamefont
  {K{\"u}fner}}\ and\ \bibinfo {author} {\bibfnamefont {F.}~\bibnamefont
  {Bechstedt}},\ }\href@noop {} {\bibfield  {journal} {\bibinfo  {journal}
  {Physical Review B}\ }\textbf {\bibinfo {volume} {89}},\ \bibinfo {pages}
  {195312} (\bibinfo {year} {2014})}\BibitemShut {NoStop}%
\bibitem [{\citenamefont {Gusev}\ \emph {et~al.}(2014)\citenamefont {Gusev},
  \citenamefont {Kvon}, \citenamefont {Olshanetsky}, \citenamefont {Levin},
  \citenamefont {Krupko}, \citenamefont {Portal}, \citenamefont {Mikhailov},\
  and\ \citenamefont {Dvoretsky}}]{8}%
  \BibitemOpen
  \bibfield  {author} {\bibinfo {author} {\bibfnamefont {G.~M.}\ \bibnamefont
  {Gusev}}, \bibinfo {author} {\bibfnamefont {Z.}~\bibnamefont {Kvon}},
  \bibinfo {author} {\bibfnamefont {E.}~\bibnamefont {Olshanetsky}}, \bibinfo
  {author} {\bibfnamefont {A.}~\bibnamefont {Levin}}, \bibinfo {author}
  {\bibfnamefont {Y.}~\bibnamefont {Krupko}}, \bibinfo {author} {\bibfnamefont
  {J.}~\bibnamefont {Portal}}, \bibinfo {author} {\bibfnamefont
  {N.}~\bibnamefont {Mikhailov}}, \ and\ \bibinfo {author} {\bibfnamefont
  {S.}~\bibnamefont {Dvoretsky}},\ }\href@noop {} {\bibfield  {journal}
  {\bibinfo  {journal} {Physical Review B}\ }\textbf {\bibinfo {volume} {89}},\
  \bibinfo {pages} {125305} (\bibinfo {year} {2014})}\BibitemShut {NoStop}%
\bibitem [{\citenamefont {Feng}\ \emph
  {et~al.}(2010{\natexlab{a}})\citenamefont {Feng}, \citenamefont {Xiao},
  \citenamefont {Zhang},\ and\ \citenamefont {Yao}}]{9}%
  \BibitemOpen
  \bibfield  {author} {\bibinfo {author} {\bibfnamefont {W.}~\bibnamefont
  {Feng}}, \bibinfo {author} {\bibfnamefont {D.}~\bibnamefont {Xiao}}, \bibinfo
  {author} {\bibfnamefont {Y.}~\bibnamefont {Zhang}}, \ and\ \bibinfo {author}
  {\bibfnamefont {Y.}~\bibnamefont {Yao}},\ }\href@noop {} {\bibfield
  {journal} {\bibinfo  {journal} {Physical Review B}\ }\textbf {\bibinfo
  {volume} {82}},\ \bibinfo {pages} {235121} (\bibinfo {year}
  {2010}{\natexlab{a}})}\BibitemShut {NoStop}%
\bibitem [{\citenamefont {Barman}\ and\ \citenamefont {Alam}(2018)}]{10}%
  \BibitemOpen
  \bibfield  {author} {\bibinfo {author} {\bibfnamefont {C.}~\bibnamefont
  {Barman}}\ and\ \bibinfo {author} {\bibfnamefont {A.}~\bibnamefont {Alam}},\
  }\href@noop {} {\bibfield  {journal} {\bibinfo  {journal} {Physical Review
  B}\ }\textbf {\bibinfo {volume} {97}},\ \bibinfo {pages} {075302} (\bibinfo
  {year} {2018})}\BibitemShut {NoStop}%
\bibitem [{\citenamefont {Shi}\ \emph {et~al.}(2017)\citenamefont {Shi},
  \citenamefont {Si}, \citenamefont {Xie}, \citenamefont {Mi}, \citenamefont
  {Xiao},\ and\ \citenamefont {Luo}}]{11}%
  \BibitemOpen
  \bibfield  {author} {\bibinfo {author} {\bibfnamefont {F.}~\bibnamefont
  {Shi}}, \bibinfo {author} {\bibfnamefont {M.}~\bibnamefont {Si}}, \bibinfo
  {author} {\bibfnamefont {J.}~\bibnamefont {Xie}}, \bibinfo {author}
  {\bibfnamefont {K.}~\bibnamefont {Mi}}, \bibinfo {author} {\bibfnamefont
  {C.}~\bibnamefont {Xiao}}, \ and\ \bibinfo {author} {\bibfnamefont
  {Q.}~\bibnamefont {Luo}},\ }\href@noop {} {\bibfield  {journal} {\bibinfo
  {journal} {Journal of Applied Physics}\ }\textbf {\bibinfo {volume} {122}},\
  \bibinfo {pages} {215701} (\bibinfo {year} {2017})}\BibitemShut {NoStop}%
\bibitem [{\citenamefont {Al-Sawai}\ \emph {et~al.}(2010)\citenamefont
  {Al-Sawai}, \citenamefont {Lin}, \citenamefont {Markiewicz}, \citenamefont
  {Wray}, \citenamefont {Xia}, \citenamefont {Xu}, \citenamefont {Hasan},\ and\
  \citenamefont {Bansil}}]{12}%
  \BibitemOpen
  \bibfield  {author} {\bibinfo {author} {\bibfnamefont {W.}~\bibnamefont
  {Al-Sawai}}, \bibinfo {author} {\bibfnamefont {H.}~\bibnamefont {Lin}},
  \bibinfo {author} {\bibfnamefont {R.}~\bibnamefont {Markiewicz}}, \bibinfo
  {author} {\bibfnamefont {L.}~\bibnamefont {Wray}}, \bibinfo {author}
  {\bibfnamefont {Y.}~\bibnamefont {Xia}}, \bibinfo {author} {\bibfnamefont
  {S.-Y.}\ \bibnamefont {Xu}}, \bibinfo {author} {\bibfnamefont
  {M.}~\bibnamefont {Hasan}}, \ and\ \bibinfo {author} {\bibfnamefont
  {A.}~\bibnamefont {Bansil}},\ }\href@noop {} {\bibfield  {journal} {\bibinfo
  {journal} {Physical Review B}\ }\textbf {\bibinfo {volume} {82}},\ \bibinfo
  {pages} {125208} (\bibinfo {year} {2010})}\BibitemShut {NoStop}%
\bibitem [{\citenamefont {Fu}\ \emph {et~al.}(2015)\citenamefont {Fu},
  \citenamefont {Bai}, \citenamefont {Liu}, \citenamefont {Tang}, \citenamefont
  {Chen}, \citenamefont {Zhao},\ and\ \citenamefont {Zhu}}]{13}%
  \BibitemOpen
  \bibfield  {author} {\bibinfo {author} {\bibfnamefont {C.}~\bibnamefont
  {Fu}}, \bibinfo {author} {\bibfnamefont {S.}~\bibnamefont {Bai}}, \bibinfo
  {author} {\bibfnamefont {Y.}~\bibnamefont {Liu}}, \bibinfo {author}
  {\bibfnamefont {Y.}~\bibnamefont {Tang}}, \bibinfo {author} {\bibfnamefont
  {L.}~\bibnamefont {Chen}}, \bibinfo {author} {\bibfnamefont {X.}~\bibnamefont
  {Zhao}}, \ and\ \bibinfo {author} {\bibfnamefont {T.}~\bibnamefont {Zhu}},\
  }\href@noop {} {\bibfield  {journal} {\bibinfo  {journal} {Nature
  communications}\ }\textbf {\bibinfo {volume} {6}},\ \bibinfo {pages} {8144}
  (\bibinfo {year} {2015})}\BibitemShut {NoStop}%
\bibitem [{\citenamefont {Zhu}\ \emph {et~al.}(2015)\citenamefont {Zhu},
  \citenamefont {Fu}, \citenamefont {Xie}, \citenamefont {Liu},\ and\
  \citenamefont {Zhao}}]{14}%
  \BibitemOpen
  \bibfield  {author} {\bibinfo {author} {\bibfnamefont {T.}~\bibnamefont
  {Zhu}}, \bibinfo {author} {\bibfnamefont {C.}~\bibnamefont {Fu}}, \bibinfo
  {author} {\bibfnamefont {H.}~\bibnamefont {Xie}}, \bibinfo {author}
  {\bibfnamefont {Y.}~\bibnamefont {Liu}}, \ and\ \bibinfo {author}
  {\bibfnamefont {X.}~\bibnamefont {Zhao}},\ }\href@noop {} {\bibfield
  {journal} {\bibinfo  {journal} {Advanced Energy Materials}\ }\textbf
  {\bibinfo {volume} {5}},\ \bibinfo {pages} {1500588} (\bibinfo {year}
  {2015})}\BibitemShut {NoStop}%
\bibitem [{\citenamefont {Xiao}\ \emph {et~al.}(2018)\citenamefont {Xiao},
  \citenamefont {Hu}, \citenamefont {Liu}, \citenamefont {Zhu}, \citenamefont
  {Li}, \citenamefont {Mu}, \citenamefont {Su}, \citenamefont {Li},\ and\
  \citenamefont {Mao}}]{15}%
  \BibitemOpen
  \bibfield  {author} {\bibinfo {author} {\bibfnamefont {H.}~\bibnamefont
  {Xiao}}, \bibinfo {author} {\bibfnamefont {T.}~\bibnamefont {Hu}}, \bibinfo
  {author} {\bibfnamefont {W.}~\bibnamefont {Liu}}, \bibinfo {author}
  {\bibfnamefont {Y.}~\bibnamefont {Zhu}}, \bibinfo {author} {\bibfnamefont
  {P.}~\bibnamefont {Li}}, \bibinfo {author} {\bibfnamefont {G.}~\bibnamefont
  {Mu}}, \bibinfo {author} {\bibfnamefont {J.}~\bibnamefont {Su}}, \bibinfo
  {author} {\bibfnamefont {K.}~\bibnamefont {Li}}, \ and\ \bibinfo {author}
  {\bibfnamefont {Z.}~\bibnamefont {Mao}},\ }\href@noop {} {\bibfield
  {journal} {\bibinfo  {journal} {Physical Review B}\ }\textbf {\bibinfo
  {volume} {97}},\ \bibinfo {pages} {224511} (\bibinfo {year}
  {2018})}\BibitemShut {NoStop}%
\bibitem [{\citenamefont {Tafti}\ \emph {et~al.}(2013)\citenamefont {Tafti},
  \citenamefont {Fujii}, \citenamefont {Juneau-Fecteau}, \citenamefont
  {de~Cotret}, \citenamefont {Doiron-Leyraud}, \citenamefont {Asamitsu},\ and\
  \citenamefont {Taillefer}}]{16}%
  \BibitemOpen
  \bibfield  {author} {\bibinfo {author} {\bibfnamefont {F.}~\bibnamefont
  {Tafti}}, \bibinfo {author} {\bibfnamefont {T.}~\bibnamefont {Fujii}},
  \bibinfo {author} {\bibfnamefont {A.}~\bibnamefont {Juneau-Fecteau}},
  \bibinfo {author} {\bibfnamefont {S.~R.}\ \bibnamefont {de~Cotret}}, \bibinfo
  {author} {\bibfnamefont {N.}~\bibnamefont {Doiron-Leyraud}}, \bibinfo
  {author} {\bibfnamefont {A.}~\bibnamefont {Asamitsu}}, \ and\ \bibinfo
  {author} {\bibfnamefont {L.}~\bibnamefont {Taillefer}},\ }\href@noop {}
  {\bibfield  {journal} {\bibinfo  {journal} {Physical Review B}\ }\textbf
  {\bibinfo {volume} {87}},\ \bibinfo {pages} {184504} (\bibinfo {year}
  {2013})}\BibitemShut {NoStop}%
\bibitem [{\citenamefont {Rozale}\ \emph {et~al.}(2013)\citenamefont {Rozale},
  \citenamefont {Amar}, \citenamefont {Lakdja}, \citenamefont {Moukadem},\ and\
  \citenamefont {Chahed}}]{17}%
  \BibitemOpen
  \bibfield  {author} {\bibinfo {author} {\bibfnamefont {H.}~\bibnamefont
  {Rozale}}, \bibinfo {author} {\bibfnamefont {A.}~\bibnamefont {Amar}},
  \bibinfo {author} {\bibfnamefont {A.}~\bibnamefont {Lakdja}}, \bibinfo
  {author} {\bibfnamefont {A.}~\bibnamefont {Moukadem}}, \ and\ \bibinfo
  {author} {\bibfnamefont {A.}~\bibnamefont {Chahed}},\ }\href@noop {}
  {\bibfield  {journal} {\bibinfo  {journal} {Journal of magnetism and magnetic
  materials}\ }\textbf {\bibinfo {volume} {336}},\ \bibinfo {pages} {83}
  (\bibinfo {year} {2013})}\BibitemShut {NoStop}%
\bibitem [{\citenamefont {Hoang}\ \emph {et~al.}(2022)\citenamefont {Hoang},
  \citenamefont {Rhim},\ and\ \citenamefont {Hong}}]{18}%
  \BibitemOpen
  \bibfield  {author} {\bibinfo {author} {\bibfnamefont {T.~T.}\ \bibnamefont
  {Hoang}}, \bibinfo {author} {\bibfnamefont {S.}~\bibnamefont {Rhim}}, \ and\
  \bibinfo {author} {\bibfnamefont {S.}~\bibnamefont {Hong}},\ }\href@noop {}
  {\bibfield  {journal} {\bibinfo  {journal} {Physical Review Materials}\
  }\textbf {\bibinfo {volume} {6}},\ \bibinfo {pages} {055001} (\bibinfo {year}
  {2022})}\BibitemShut {NoStop}%
\bibitem [{\citenamefont {Casper}\ \emph {et~al.}(2012)\citenamefont {Casper},
  \citenamefont {Graf}, \citenamefont {Chadov}, \citenamefont {Balke},\ and\
  \citenamefont {Felser}}]{31}%
  \BibitemOpen
  \bibfield  {author} {\bibinfo {author} {\bibfnamefont {F.}~\bibnamefont
  {Casper}}, \bibinfo {author} {\bibfnamefont {T.}~\bibnamefont {Graf}},
  \bibinfo {author} {\bibfnamefont {S.}~\bibnamefont {Chadov}}, \bibinfo
  {author} {\bibfnamefont {B.}~\bibnamefont {Balke}}, \ and\ \bibinfo {author}
  {\bibfnamefont {C.}~\bibnamefont {Felser}},\ }\href@noop {} {\bibfield
  {journal} {\bibinfo  {journal} {Semiconductor Science and Technology}\
  }\textbf {\bibinfo {volume} {27}},\ \bibinfo {pages} {063001} (\bibinfo
  {year} {2012})}\BibitemShut {NoStop}%
\bibitem [{\citenamefont {Shekhar}\ \emph {et~al.}(2018)\citenamefont
  {Shekhar}, \citenamefont {Kumar}, \citenamefont {Grinenko}, \citenamefont
  {Singh}, \citenamefont {Sarkar}, \citenamefont {Luetkens}, \citenamefont
  {Wu}, \citenamefont {Zhang}, \citenamefont {Komarek}, \citenamefont {Kampert}
  \emph {et~al.}}]{32}%
  \BibitemOpen
  \bibfield  {author} {\bibinfo {author} {\bibfnamefont {C.}~\bibnamefont
  {Shekhar}}, \bibinfo {author} {\bibfnamefont {N.}~\bibnamefont {Kumar}},
  \bibinfo {author} {\bibfnamefont {V.}~\bibnamefont {Grinenko}}, \bibinfo
  {author} {\bibfnamefont {S.}~\bibnamefont {Singh}}, \bibinfo {author}
  {\bibfnamefont {R.}~\bibnamefont {Sarkar}}, \bibinfo {author} {\bibfnamefont
  {H.}~\bibnamefont {Luetkens}}, \bibinfo {author} {\bibfnamefont {S.-C.}\
  \bibnamefont {Wu}}, \bibinfo {author} {\bibfnamefont {Y.}~\bibnamefont
  {Zhang}}, \bibinfo {author} {\bibfnamefont {A.~C.}\ \bibnamefont {Komarek}},
  \bibinfo {author} {\bibfnamefont {E.}~\bibnamefont {Kampert}},  \emph
  {et~al.},\ }\href@noop {} {\bibfield  {journal} {\bibinfo  {journal}
  {Proceedings of the National Academy of Sciences}\ }\textbf {\bibinfo
  {volume} {115}},\ \bibinfo {pages} {9140} (\bibinfo {year}
  {2018})}\BibitemShut {NoStop}%
\bibitem [{\citenamefont {Sukhanov}\ \emph {et~al.}(2020)\citenamefont
  {Sukhanov}, \citenamefont {Onykiienko}, \citenamefont {Bewley}, \citenamefont
  {Shekhar}, \citenamefont {Felser},\ and\ \citenamefont {Inosov}}]{33}%
  \BibitemOpen
  \bibfield  {author} {\bibinfo {author} {\bibfnamefont {A.}~\bibnamefont
  {Sukhanov}}, \bibinfo {author} {\bibfnamefont {Y.}~\bibnamefont
  {Onykiienko}}, \bibinfo {author} {\bibfnamefont {R.}~\bibnamefont {Bewley}},
  \bibinfo {author} {\bibfnamefont {C.}~\bibnamefont {Shekhar}}, \bibinfo
  {author} {\bibfnamefont {C.}~\bibnamefont {Felser}}, \ and\ \bibinfo {author}
  {\bibfnamefont {D.}~\bibnamefont {Inosov}},\ }\href@noop {} {\bibfield
  {journal} {\bibinfo  {journal} {Physical Review B}\ }\textbf {\bibinfo
  {volume} {101}},\ \bibinfo {pages} {014417} (\bibinfo {year}
  {2020})}\BibitemShut {NoStop}%
\bibitem [{\citenamefont {Kieven}\ \emph {et~al.}(2010)\citenamefont {Kieven},
  \citenamefont {Klenk}, \citenamefont {Naghavi}, \citenamefont {Felser},\ and\
  \citenamefont {Gruhn}}]{23}%
  \BibitemOpen
  \bibfield  {author} {\bibinfo {author} {\bibfnamefont {D.}~\bibnamefont
  {Kieven}}, \bibinfo {author} {\bibfnamefont {R.}~\bibnamefont {Klenk}},
  \bibinfo {author} {\bibfnamefont {S.}~\bibnamefont {Naghavi}}, \bibinfo
  {author} {\bibfnamefont {C.}~\bibnamefont {Felser}}, \ and\ \bibinfo {author}
  {\bibfnamefont {T.}~\bibnamefont {Gruhn}},\ }\href@noop {} {\bibfield
  {journal} {\bibinfo  {journal} {Physical Review B}\ }\textbf {\bibinfo
  {volume} {81}},\ \bibinfo {pages} {075208} (\bibinfo {year}
  {2010})}\BibitemShut {NoStop}%
\bibitem [{\citenamefont {Abada}\ and\ \citenamefont {Marbouh}(2020)}]{24}%
  \BibitemOpen
  \bibfield  {author} {\bibinfo {author} {\bibfnamefont {A.}~\bibnamefont
  {Abada}}\ and\ \bibinfo {author} {\bibfnamefont {N.}~\bibnamefont
  {Marbouh}},\ }\href@noop {} {\bibfield  {journal} {\bibinfo  {journal}
  {Journal of Superconductivity and Novel Magnetism}\ }\textbf {\bibinfo
  {volume} {33}},\ \bibinfo {pages} {889} (\bibinfo {year} {2020})}\BibitemShut
  {NoStop}%
\bibitem [{\citenamefont {Rached}\ \emph {et~al.}(2022)\citenamefont {Rached},
  \citenamefont {Caid}, \citenamefont {Rached}, \citenamefont {Merabet},
  \citenamefont {Benalia}, \citenamefont {Al-Qaisi}, \citenamefont {Djoudi},\
  and\ \citenamefont {Rached}}]{25}%
  \BibitemOpen
  \bibfield  {author} {\bibinfo {author} {\bibfnamefont {Y.}~\bibnamefont
  {Rached}}, \bibinfo {author} {\bibfnamefont {M.}~\bibnamefont {Caid}},
  \bibinfo {author} {\bibfnamefont {H.}~\bibnamefont {Rached}}, \bibinfo
  {author} {\bibfnamefont {M.}~\bibnamefont {Merabet}}, \bibinfo {author}
  {\bibfnamefont {S.}~\bibnamefont {Benalia}}, \bibinfo {author} {\bibfnamefont
  {S.}~\bibnamefont {Al-Qaisi}}, \bibinfo {author} {\bibfnamefont
  {L.}~\bibnamefont {Djoudi}}, \ and\ \bibinfo {author} {\bibfnamefont
  {D.}~\bibnamefont {Rached}},\ }\href@noop {} {\bibfield  {journal} {\bibinfo
  {journal} {Journal of Superconductivity and Novel Magnetism}\ }\textbf
  {\bibinfo {volume} {35}},\ \bibinfo {pages} {875} (\bibinfo {year}
  {2022})}\BibitemShut {NoStop}%
\bibitem [{\citenamefont {Galanakis}\ \emph {et~al.}(2002)\citenamefont
  {Galanakis}, \citenamefont {Dederichs},\ and\ \citenamefont
  {Papanikolaou}}]{19}%
  \BibitemOpen
  \bibfield  {author} {\bibinfo {author} {\bibfnamefont {I.}~\bibnamefont
  {Galanakis}}, \bibinfo {author} {\bibfnamefont {P.}~\bibnamefont
  {Dederichs}}, \ and\ \bibinfo {author} {\bibfnamefont {N.}~\bibnamefont
  {Papanikolaou}},\ }\href@noop {} {\bibfield  {journal} {\bibinfo  {journal}
  {Physical Review B}\ }\textbf {\bibinfo {volume} {66}},\ \bibinfo {pages}
  {174429} (\bibinfo {year} {2002})}\BibitemShut {NoStop}%
\bibitem [{\citenamefont {Lin}\ \emph {et~al.}(2015)\citenamefont {Lin},
  \citenamefont {Chen}, \citenamefont {Yang}, \citenamefont {Zhao},
  \citenamefont {Wu}, \citenamefont {Felser},\ and\ \citenamefont {Yan}}]{20}%
  \BibitemOpen
  \bibfield  {author} {\bibinfo {author} {\bibfnamefont {S.-Y.}\ \bibnamefont
  {Lin}}, \bibinfo {author} {\bibfnamefont {M.}~\bibnamefont {Chen}}, \bibinfo
  {author} {\bibfnamefont {X.-B.}\ \bibnamefont {Yang}}, \bibinfo {author}
  {\bibfnamefont {Y.-J.}\ \bibnamefont {Zhao}}, \bibinfo {author}
  {\bibfnamefont {S.-C.}\ \bibnamefont {Wu}}, \bibinfo {author} {\bibfnamefont
  {C.}~\bibnamefont {Felser}}, \ and\ \bibinfo {author} {\bibfnamefont
  {B.}~\bibnamefont {Yan}},\ }\href@noop {} {\bibfield  {journal} {\bibinfo
  {journal} {Physical Review B}\ }\textbf {\bibinfo {volume} {91}},\ \bibinfo
  {pages} {094107} (\bibinfo {year} {2015})}\BibitemShut {NoStop}%
\bibitem [{\citenamefont {Xiao}\ \emph
  {et~al.}(2010{\natexlab{a}})\citenamefont {Xiao}, \citenamefont {Yao},
  \citenamefont {Feng}, \citenamefont {Wen}, \citenamefont {Zhu}, \citenamefont
  {Chen}, \citenamefont {Stocks},\ and\ \citenamefont {Zhang}}]{21}%
  \BibitemOpen
  \bibfield  {author} {\bibinfo {author} {\bibfnamefont {D.}~\bibnamefont
  {Xiao}}, \bibinfo {author} {\bibfnamefont {Y.}~\bibnamefont {Yao}}, \bibinfo
  {author} {\bibfnamefont {W.}~\bibnamefont {Feng}}, \bibinfo {author}
  {\bibfnamefont {J.}~\bibnamefont {Wen}}, \bibinfo {author} {\bibfnamefont
  {W.}~\bibnamefont {Zhu}}, \bibinfo {author} {\bibfnamefont {X.-Q.}\
  \bibnamefont {Chen}}, \bibinfo {author} {\bibfnamefont {G.~M.}\ \bibnamefont
  {Stocks}}, \ and\ \bibinfo {author} {\bibfnamefont {Z.}~\bibnamefont
  {Zhang}},\ }\href@noop {} {\bibfield  {journal} {\bibinfo  {journal}
  {Physical review letters}\ }\textbf {\bibinfo {volume} {105}},\ \bibinfo
  {pages} {096404} (\bibinfo {year} {2010}{\natexlab{a}})}\BibitemShut
  {NoStop}%
\bibitem [{\citenamefont {Gofryk}\ \emph {et~al.}(2011)\citenamefont {Gofryk},
  \citenamefont {Kaczorowski}, \citenamefont {Plackowski}, \citenamefont
  {Leithe-Jasper},\ and\ \citenamefont {Grin}}]{34}%
  \BibitemOpen
  \bibfield  {author} {\bibinfo {author} {\bibfnamefont {K.}~\bibnamefont
  {Gofryk}}, \bibinfo {author} {\bibfnamefont {D.}~\bibnamefont {Kaczorowski}},
  \bibinfo {author} {\bibfnamefont {T.}~\bibnamefont {Plackowski}}, \bibinfo
  {author} {\bibfnamefont {A.}~\bibnamefont {Leithe-Jasper}}, \ and\ \bibinfo
  {author} {\bibfnamefont {Y.}~\bibnamefont {Grin}},\ }\href@noop {} {\bibfield
   {journal} {\bibinfo  {journal} {Physical Review B}\ }\textbf {\bibinfo
  {volume} {84}},\ \bibinfo {pages} {035208} (\bibinfo {year}
  {2011})}\BibitemShut {NoStop}%
\bibitem [{\citenamefont {Dhori}\ \emph {et~al.}(2022)\citenamefont {Dhori},
  \citenamefont {Sattigeri}, \citenamefont {Jha}, \citenamefont
  {Kurzydlowski},\ and\ \citenamefont {Chakraborty}}]{28}%
  \BibitemOpen
  \bibfield  {author} {\bibinfo {author} {\bibfnamefont {B.~R.}\ \bibnamefont
  {Dhori}}, \bibinfo {author} {\bibfnamefont {R.~M.}\ \bibnamefont
  {Sattigeri}}, \bibinfo {author} {\bibfnamefont {P.~K.}\ \bibnamefont {Jha}},
  \bibinfo {author} {\bibfnamefont {D.}~\bibnamefont {Kurzydlowski}}, \ and\
  \bibinfo {author} {\bibfnamefont {B.}~\bibnamefont {Chakraborty}},\
  }\href@noop {} {\bibfield  {journal} {\bibinfo  {journal} {Materials
  Advances}\ }\textbf {\bibinfo {volume} {3}},\ \bibinfo {pages} {3938}
  (\bibinfo {year} {2022})}\BibitemShut {NoStop}%
\bibitem [{\citenamefont {Br{\"u}ne}\ \emph {et~al.}(2011)\citenamefont
  {Br{\"u}ne}, \citenamefont {Liu}, \citenamefont {Novik}, \citenamefont
  {Hankiewicz}, \citenamefont {Buhmann}, \citenamefont {Chen}, \citenamefont
  {Qi}, \citenamefont {Shen}, \citenamefont {Zhang},\ and\ \citenamefont
  {Molenkamp}}]{22}%
  \BibitemOpen
  \bibfield  {author} {\bibinfo {author} {\bibfnamefont {C.}~\bibnamefont
  {Br{\"u}ne}}, \bibinfo {author} {\bibfnamefont {C.}~\bibnamefont {Liu}},
  \bibinfo {author} {\bibfnamefont {E.}~\bibnamefont {Novik}}, \bibinfo
  {author} {\bibfnamefont {E.}~\bibnamefont {Hankiewicz}}, \bibinfo {author}
  {\bibfnamefont {H.}~\bibnamefont {Buhmann}}, \bibinfo {author} {\bibfnamefont
  {Y.}~\bibnamefont {Chen}}, \bibinfo {author} {\bibfnamefont {X.}~\bibnamefont
  {Qi}}, \bibinfo {author} {\bibfnamefont {Z.}~\bibnamefont {Shen}}, \bibinfo
  {author} {\bibfnamefont {S.}~\bibnamefont {Zhang}}, \ and\ \bibinfo {author}
  {\bibfnamefont {L.}~\bibnamefont {Molenkamp}},\ }\href@noop {} {\bibfield
  {journal} {\bibinfo  {journal} {Physical Review Letters}\ }\textbf {\bibinfo
  {volume} {106}},\ \bibinfo {pages} {126803} (\bibinfo {year}
  {2011})}\BibitemShut {NoStop}%
\bibitem [{\citenamefont {Weidenkaff}\ \emph {et~al.}(1999)\citenamefont
  {Weidenkaff}, \citenamefont {Steinfeld}, \citenamefont {Wokaun},
  \citenamefont {Auer}, \citenamefont {Eichler},\ and\ \citenamefont
  {Reller}}]{r6}%
  \BibitemOpen
  \bibfield  {author} {\bibinfo {author} {\bibfnamefont {A.}~\bibnamefont
  {Weidenkaff}}, \bibinfo {author} {\bibfnamefont {A.}~\bibnamefont
  {Steinfeld}}, \bibinfo {author} {\bibfnamefont {A.}~\bibnamefont {Wokaun}},
  \bibinfo {author} {\bibfnamefont {P.}~\bibnamefont {Auer}}, \bibinfo {author}
  {\bibfnamefont {B.}~\bibnamefont {Eichler}}, \ and\ \bibinfo {author}
  {\bibfnamefont {A.}~\bibnamefont {Reller}},\ }\href@noop {} {\bibfield
  {journal} {\bibinfo  {journal} {Solar energy}\ }\textbf {\bibinfo {volume}
  {65}},\ \bibinfo {pages} {59} (\bibinfo {year} {1999})}\BibitemShut {NoStop}%
\bibitem [{\citenamefont {Shrivastava}\ and\ \citenamefont
  {Sanyal}(2018)}]{29}%
  \BibitemOpen
  \bibfield  {author} {\bibinfo {author} {\bibfnamefont {D.}~\bibnamefont
  {Shrivastava}}\ and\ \bibinfo {author} {\bibfnamefont {S.~P.}\ \bibnamefont
  {Sanyal}},\ }\href@noop {} {\bibfield  {journal} {\bibinfo  {journal} {Solid
  State Communications}\ }\textbf {\bibinfo {volume} {273}},\ \bibinfo {pages}
  {1} (\bibinfo {year} {2018})}\BibitemShut {NoStop}%
\bibitem [{\citenamefont {Giannozzi}\ \emph {et~al.}(2017)\citenamefont
  {Giannozzi}, \citenamefont {Andreussi}, \citenamefont {Brumme}, \citenamefont
  {Bunau}, \citenamefont {Nardelli}, \citenamefont {Calandra}, \citenamefont
  {Car}, \citenamefont {Cavazzoni}, \citenamefont {Ceresoli}, \citenamefont
  {Cococcioni} \emph {et~al.}}]{qe}%
  \BibitemOpen
  \bibfield  {author} {\bibinfo {author} {\bibfnamefont {P.}~\bibnamefont
  {Giannozzi}}, \bibinfo {author} {\bibfnamefont {O.}~\bibnamefont
  {Andreussi}}, \bibinfo {author} {\bibfnamefont {T.}~\bibnamefont {Brumme}},
  \bibinfo {author} {\bibfnamefont {O.}~\bibnamefont {Bunau}}, \bibinfo
  {author} {\bibfnamefont {M.~B.}\ \bibnamefont {Nardelli}}, \bibinfo {author}
  {\bibfnamefont {M.}~\bibnamefont {Calandra}}, \bibinfo {author}
  {\bibfnamefont {R.}~\bibnamefont {Car}}, \bibinfo {author} {\bibfnamefont
  {C.}~\bibnamefont {Cavazzoni}}, \bibinfo {author} {\bibfnamefont
  {D.}~\bibnamefont {Ceresoli}}, \bibinfo {author} {\bibfnamefont
  {M.}~\bibnamefont {Cococcioni}},  \emph {et~al.},\ }\href@noop {} {\bibfield
  {journal} {\bibinfo  {journal} {Journal of physics: Condensed matter}\
  }\textbf {\bibinfo {volume} {29}},\ \bibinfo {pages} {465901} (\bibinfo
  {year} {2017})}\BibitemShut {NoStop}%
\bibitem [{\citenamefont {Perdew}\ \emph {et~al.}(1996)\citenamefont {Perdew},
  \citenamefont {Burke},\ and\ \citenamefont {Ernzerhof}}]{35}%
  \BibitemOpen
  \bibfield  {author} {\bibinfo {author} {\bibfnamefont {J.~P.}\ \bibnamefont
  {Perdew}}, \bibinfo {author} {\bibfnamefont {K.}~\bibnamefont {Burke}}, \
  and\ \bibinfo {author} {\bibfnamefont {M.}~\bibnamefont {Ernzerhof}},\
  }\href@noop {} {\bibfield  {journal} {\bibinfo  {journal} {Physical review
  letters}\ }\textbf {\bibinfo {volume} {77}},\ \bibinfo {pages} {3865}
  (\bibinfo {year} {1996})}\BibitemShut {NoStop}%
\bibitem [{\citenamefont {Monkhorst}\ and\ \citenamefont {Pack}(1976)}]{36}%
  \BibitemOpen
  \bibfield  {author} {\bibinfo {author} {\bibfnamefont {H.~J.}\ \bibnamefont
  {Monkhorst}}\ and\ \bibinfo {author} {\bibfnamefont {J.~D.}\ \bibnamefont
  {Pack}},\ }\href@noop {} {\bibfield  {journal} {\bibinfo  {journal} {Physical
  review B}\ }\textbf {\bibinfo {volume} {13}},\ \bibinfo {pages} {5188}
  (\bibinfo {year} {1976})}\BibitemShut {NoStop}%
\bibitem [{\citenamefont {Marzari}\ \emph {et~al.}(2012)\citenamefont
  {Marzari}, \citenamefont {Mostofi}, \citenamefont {Yates}, \citenamefont
  {Souza},\ and\ \citenamefont {Vanderbilt}}]{37}%
  \BibitemOpen
  \bibfield  {author} {\bibinfo {author} {\bibfnamefont {N.}~\bibnamefont
  {Marzari}}, \bibinfo {author} {\bibfnamefont {A.~A.}\ \bibnamefont
  {Mostofi}}, \bibinfo {author} {\bibfnamefont {J.~R.}\ \bibnamefont {Yates}},
  \bibinfo {author} {\bibfnamefont {I.}~\bibnamefont {Souza}}, \ and\ \bibinfo
  {author} {\bibfnamefont {D.}~\bibnamefont {Vanderbilt}},\ }\href@noop {}
  {\bibfield  {journal} {\bibinfo  {journal} {Reviews of Modern Physics}\
  }\textbf {\bibinfo {volume} {84}},\ \bibinfo {pages} {1419} (\bibinfo {year}
  {2012})}\BibitemShut {NoStop}%
\bibitem [{\citenamefont {Mostofi}\ \emph {et~al.}(2008)\citenamefont
  {Mostofi}, \citenamefont {Yates}, \citenamefont {Lee}, \citenamefont {Souza},
  \citenamefont {Vanderbilt},\ and\ \citenamefont {Marzari}}]{38}%
  \BibitemOpen
  \bibfield  {author} {\bibinfo {author} {\bibfnamefont {A.~A.}\ \bibnamefont
  {Mostofi}}, \bibinfo {author} {\bibfnamefont {J.~R.}\ \bibnamefont {Yates}},
  \bibinfo {author} {\bibfnamefont {Y.-S.}\ \bibnamefont {Lee}}, \bibinfo
  {author} {\bibfnamefont {I.}~\bibnamefont {Souza}}, \bibinfo {author}
  {\bibfnamefont {D.}~\bibnamefont {Vanderbilt}}, \ and\ \bibinfo {author}
  {\bibfnamefont {N.}~\bibnamefont {Marzari}},\ }\href@noop {} {\bibfield
  {journal} {\bibinfo  {journal} {Computer physics communications}\ }\textbf
  {\bibinfo {volume} {178}},\ \bibinfo {pages} {685} (\bibinfo {year}
  {2008})}\BibitemShut {NoStop}%
\bibitem [{\citenamefont {Wu}\ \emph {et~al.}(2018)\citenamefont {Wu},
  \citenamefont {Zhang}, \citenamefont {Song}, \citenamefont {Troyer},\ and\
  \citenamefont {Soluyanov}}]{39}%
  \BibitemOpen
  \bibfield  {author} {\bibinfo {author} {\bibfnamefont {Q.}~\bibnamefont
  {Wu}}, \bibinfo {author} {\bibfnamefont {S.}~\bibnamefont {Zhang}}, \bibinfo
  {author} {\bibfnamefont {H.-F.}\ \bibnamefont {Song}}, \bibinfo {author}
  {\bibfnamefont {M.}~\bibnamefont {Troyer}}, \ and\ \bibinfo {author}
  {\bibfnamefont {A.~A.}\ \bibnamefont {Soluyanov}},\ }\href@noop {} {\bibfield
   {journal} {\bibinfo  {journal} {Computer Physics Communications}\ }\textbf
  {\bibinfo {volume} {224}},\ \bibinfo {pages} {405} (\bibinfo {year}
  {2018})}\BibitemShut {NoStop}%
\bibitem [{\citenamefont {Gruhn}(2010)}]{r1}%
  \BibitemOpen
  \bibfield  {author} {\bibinfo {author} {\bibfnamefont {T.}~\bibnamefont
  {Gruhn}},\ }\href@noop {} {\bibfield  {journal} {\bibinfo  {journal}
  {Physical Review B}\ }\textbf {\bibinfo {volume} {82}},\ \bibinfo {pages}
  {125210} (\bibinfo {year} {2010})}\BibitemShut {NoStop}%
\bibitem [{\citenamefont {Xiao}\ \emph
  {et~al.}(2010{\natexlab{b}})\citenamefont {Xiao}, \citenamefont {Yao},
  \citenamefont {Feng}, \citenamefont {Wen}, \citenamefont {Zhu}, \citenamefont
  {Chen}, \citenamefont {Stocks},\ and\ \citenamefont {Zhang}}]{r2}%
  \BibitemOpen
  \bibfield  {author} {\bibinfo {author} {\bibfnamefont {D.}~\bibnamefont
  {Xiao}}, \bibinfo {author} {\bibfnamefont {Y.}~\bibnamefont {Yao}}, \bibinfo
  {author} {\bibfnamefont {W.}~\bibnamefont {Feng}}, \bibinfo {author}
  {\bibfnamefont {J.}~\bibnamefont {Wen}}, \bibinfo {author} {\bibfnamefont
  {W.}~\bibnamefont {Zhu}}, \bibinfo {author} {\bibfnamefont {X.-Q.}\
  \bibnamefont {Chen}}, \bibinfo {author} {\bibfnamefont {G.~M.}\ \bibnamefont
  {Stocks}}, \ and\ \bibinfo {author} {\bibfnamefont {Z.}~\bibnamefont
  {Zhang}},\ }\href@noop {} {\bibfield  {journal} {\bibinfo  {journal}
  {Physical review letters}\ }\textbf {\bibinfo {volume} {105}},\ \bibinfo
  {pages} {096404} (\bibinfo {year} {2010}{\natexlab{b}})}\BibitemShut
  {NoStop}%
\bibitem [{\citenamefont {Singh}\ \emph {et~al.}(2020)\citenamefont {Singh},
  \citenamefont {Ramarao},\ and\ \citenamefont {Peter}}]{r3}%
  \BibitemOpen
  \bibfield  {author} {\bibinfo {author} {\bibfnamefont {A.~K.}\ \bibnamefont
  {Singh}}, \bibinfo {author} {\bibfnamefont {S.}~\bibnamefont {Ramarao}}, \
  and\ \bibinfo {author} {\bibfnamefont {S.~C.}\ \bibnamefont {Peter}},\
  }\href@noop {} {\bibfield  {journal} {\bibinfo  {journal} {APL Materials}\
  }\textbf {\bibinfo {volume} {8}},\ \bibinfo {pages} {060903} (\bibinfo {year}
  {2020})}\BibitemShut {NoStop}%
\bibitem [{\citenamefont {Kore}\ \emph {et~al.}(2022)\citenamefont {Kore},
  \citenamefont {Ara},\ and\ \citenamefont {Singh}}]{r4}%
  \BibitemOpen
  \bibfield  {author} {\bibinfo {author} {\bibfnamefont {A.}~\bibnamefont
  {Kore}}, \bibinfo {author} {\bibfnamefont {N.}~\bibnamefont {Ara}}, \ and\
  \bibinfo {author} {\bibfnamefont {P.}~\bibnamefont {Singh}},\ }\href@noop {}
  {\bibfield  {journal} {\bibinfo  {journal} {Journal of Physics: Condensed
  Matter}\ }\textbf {\bibinfo {volume} {34}},\ \bibinfo {pages} {205501}
  (\bibinfo {year} {2022})}\BibitemShut {NoStop}%
\bibitem [{\citenamefont {Feng}\ \emph
  {et~al.}(2010{\natexlab{b}})\citenamefont {Feng}, \citenamefont {Xiao},
  \citenamefont {Zhang},\ and\ \citenamefont {Yao}}]{r7}%
  \BibitemOpen
  \bibfield  {author} {\bibinfo {author} {\bibfnamefont {W.}~\bibnamefont
  {Feng}}, \bibinfo {author} {\bibfnamefont {D.}~\bibnamefont {Xiao}}, \bibinfo
  {author} {\bibfnamefont {Y.}~\bibnamefont {Zhang}}, \ and\ \bibinfo {author}
  {\bibfnamefont {Y.}~\bibnamefont {Yao}},\ }\href@noop {} {\bibfield
  {journal} {\bibinfo  {journal} {Physical Review B}\ }\textbf {\bibinfo
  {volume} {82}},\ \bibinfo {pages} {235121} (\bibinfo {year}
  {2010}{\natexlab{b}})}\BibitemShut {NoStop}%
\bibitem [{\citenamefont {Fu}\ \emph {et~al.}(2007)\citenamefont {Fu},
  \citenamefont {Kane},\ and\ \citenamefont {Mele}}]{r5}%
  \BibitemOpen
  \bibfield  {author} {\bibinfo {author} {\bibfnamefont {L.}~\bibnamefont
  {Fu}}, \bibinfo {author} {\bibfnamefont {C.~L.}\ \bibnamefont {Kane}}, \ and\
  \bibinfo {author} {\bibfnamefont {E.~J.}\ \bibnamefont {Mele}},\ }\href@noop
  {} {\bibfield  {journal} {\bibinfo  {journal} {Physical review letters}\
  }\textbf {\bibinfo {volume} {98}},\ \bibinfo {pages} {106803} (\bibinfo
  {year} {2007})}\BibitemShut {NoStop}%
\end{thebibliography}%

\end{document}